\documentclass[10pt]{iopart}

\usepackage{iopams}
\usepackage{graphicx} 
\usepackage{dcolumn}
\usepackage{times}
\usepackage{textcomp}
\usepackage{titlesec}
\usepackage{float}
\usepackage{color,xcolor}
\usepackage{graphicx,times} 
\usepackage{cuted} \stripsep -3pt plus 3pt minus 2pt
\usepackage{epstopdf}
\usepackage{graphicx}
\usepackage{subfigure}
\usepackage{makecell}
\usepackage{multirow}
\renewcommand{\figurename}{Figure}
\usepackage[colorlinks,linkcolor=blue, urlcolor=blue, anchorcolor=blue, citecolor=blue]{hyperref}
\usepackage{longtable,booktabs}
\usepackage{fouriernc}
\usepackage{epsfig,graphics,graphicx}
\usepackage{bm}
\usepackage{color}
\usepackage{flushend}
\usepackage{ulem}
\usepackage[sort&compress,square,numbers]{natbib}

\begin{document}
\title{Bremsstrahlung Radiation Power in Fusion Plasmas Revisited: Towards Accurate Analytical Fitting}

\author{
Huasheng Xie$^{1,2}$
}

\address{

$^1$ Hebei Key Laboratory of Compact Fusion, Langfang 065001, People's Republic of China

$^2$ ENN Science and Technology Development Co., Ltd., Langfang 065001, People's Republic of China

}
\eads{\mailto{huashengxie@gmail.com, xiehuasheng@enn.cn}}

\begin{indented}
\item[\today]
\end{indented}

\begin{abstract}
In fusion plasmas, where electron temperatures $T_e$ range from keV to hundreds of keV, Bremsstrahlung radiation constitutes a significant energy loss mechanism. While various thermal average fitting formulas exist in the literature, their accuracy is limited, particularly for $T_e \leq 20$ keV with error $>10\%$. Additionally, non-relativistic fitting formulas become invalid for $T_e \geq 50$ keV. The accurate calculation of Bremsstrahlung radiation is important for fusion gain study, especially of advanced fuels fusion. In this work, we develop new but still simple fitting formulas that are valid for electron temperatures ranging from small ($\lesssim0.1$ keV) to extreme relativistic range, with errors of less than 1\% even for high atomic number ions (e.g., $Z=30$), which could be sufficient for fusion plasma applications. Both electron-ion (e-i) and electron-electron (e-e) Bremsstrahlung radiations are considered. Both polynomial fitting with truncated ($t=k_BT_e/m_ec^2\leq10$) and one-fit-all formulas are provided. Additionally, we offer fitting formulas for e-i and e-e specifically for electron energies, which could prove useful in non-Maxwellian Bremsstrahlung radiation studies.
\end{abstract}
\maketitle
\ioptwocol

\section{Introduction}\label{sec:intro}

Bremsstrahlung radiation \cite{Spitzer1956,Bekefi1966} is electromagnetic radiation produced by the deceleration of a charged particle when colliding with another charged particle. The incoming particle loses kinetic energy, which is converted into radiation, i.e., photons. In general, we refer to bremsstrahlung as the radiation emitted by an electron as it decelerates in the field of other charged particles. In fusion plasmas, since the atoms are usually fully ionized, the major Bremsstrahlung radiation loss is electron-ion (e-i) radiation, i.e., free-free radiation. The e-i radiation formula with a simplified Gaunt factor can be found in textbooks \cite{Spitzer1956,Glasstone1960,Gross1984}. The widely used plasma radiation monograph by Bekefi \cite{Bekefi1966} also provides a comprehensive description of this process. When the temperature is higher than tens of keV, electron-electron (e-e) radiation would also be significant. In this work, we are interested in the total Bremsstrahlung power of these two types, and would like to obtain it accurately with an error of less than 1\%, which would then be sufficient for applications. We seek simple but accurate analytical fitting formulas, which are not only important for advanced fuels { [such as D-D (Deuterium), D-$^{3}$He (Helium), p-$^{11}$B (proton–Boron)]} \cite{Dawson1981,Rider1997,Nevins1998,Putvinski2019,Cai2022,Xie2023,Liu2024} fusion with temperatures larger than 50 keV, but also useful for D-T {(Trillium)} fusion at temperatures less than 30 keV.

\begin{figure*}
\centering
\includegraphics[width=17.5cm]{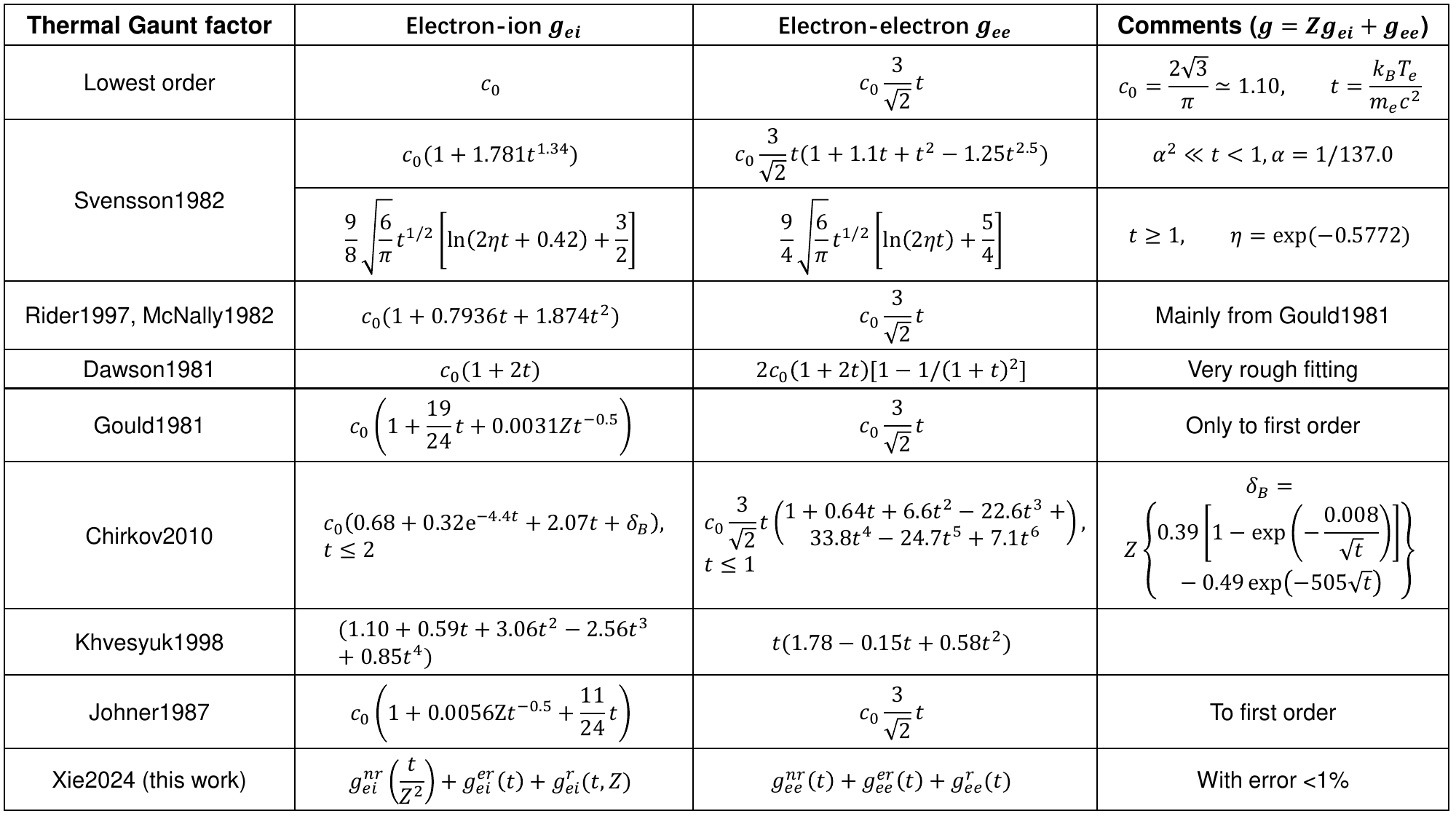}
\caption{Several usual used Bremsstrahlung radiation thermal average Gaunt factor fitting formulas in the fusion plasma community. The total Gaunt factor is $g=Zg_{ei}+g_{ee}$. { The purpose of the present work (Xie2024) is to enhance the fitting to achieve an error of less than 1\%. } }\label{fig:existfitting}
\end{figure*}

Revisiting this problem, we found that Bekefi's work on plasma radiation and the review by Koch and Motz \cite{Koch1959} did not include the later-developed relativistic models of e-i and e-e radiation, which could be important for temperatures high up to around 100 keV. Additionally, while many authors \cite{Svensson1982,McNally1982,Rider1997,Johner1987,Gould1980,Chirkov2010,Khvesyuk1998,Khvesyuk2000,Dawson1981,Stepney1983} have provided widely-used fitting formulas, their valid ranges are limited, and the errors can be large when used without caution. For $Z \neq 1$, the fitting formulas are usually even less accurate. However, we found that the most accurate yet still simple fitting is provided by Svensson \cite{Svensson1982} and used by Stepney \cite{Stepney1983} for both e-i and e-e radiation, which can still be improved for low temperature and high $Z$. Chirkov et al. \cite{Chirkov2010,Khvesyuk1998} have provided new fitting to be more accurate at low temperatures; however, the accuracy is still found not to be better than Svensson's in some regions, say at around 5\%. Gould \cite{Gould1980} only considers the lowest relativistic effect and thus is not valid for high temperatures. Gould's fitting is partly used by McNally \cite{McNally1982} and Rider \cite{Rider1997}, and the error can be larger than 30\% for $T_e > 100$ keV. Dawson's \cite{Dawson1981} fitting also oversimplifies. Care has been taken to improve the accuracy of the fitting, especially to account for the ignored effects in some simple models. The usually used fitting formulas in the fusion plasma community are listed in Figure \ref{fig:existfitting}. There are also some other fittings, see for example \cite{Brussaard1962,Stickforth1961}, which are less used in the fusion plasma community.

In astrophysics, Bremsstrahlung radiation is calculated with high accuracy, reaching below $0.1\%$ in most ranges recently, as demonstrated by van Hoof et al. \cite{Hoof2014,Hoof2015}, Chluba et al. \cite{Chluba2020} for e-i radiation, and by Nozawa et al. \cite{Nozawa2009} for e-e radiation. They provide not only comprehensive calculations of the power but also the spectrum. The most accurate e-i models are provided by Elwert and Haug (EH) \cite{Elwert1969}, and the most accurate e-e model is provided by Haug \cite{Haug1975a,Haug1975b}, valid for both non-relativistic and relativistic ranges. Seltzer and Berger \cite{Seltzer1986} provide a comprehensive set of tabulated Bremsstrahlung cross sections (differential in the energy of the emitted photons) with energy spectra from electrons with kinetic energy from 1 keV to 10 GeV incident on screened nuclei and orbital electrons of neutral atoms with $Z = 1-100$. However, here we are only interested in fully ionized plasmas.

The accurate calculation of non-relativistic e-i power radiated from an ionized gas as a result of Bremsstrahlung is based on Berger's\cite{Berger1957} evaluation of Sommerfeld's theory, which neglects only relativistic and retardation effects. Sommerfeld \cite{Sommerfeld1939}, using exact Coulomb wave functions derived from the non-relativistic Schrödinger equation, evaluated the dipole matrix elements for radiation by a free electron in the field of an ion. The results have been obtained by Berger \cite{Berger1957}, Greene \cite{Greene1959}, Karzas and Latter \cite{Karzas1961}, and repeated by van Hoof \cite{Hoof2014}. The relativistic Born approximation for e-i radiation, which is not valid for low temperatures, has also been repeated \cite{Itoh1985,Nozawa1998,Hoof2015,Chluba2020}. The non-relativistic e-e radiation is provided by Maxon and Corman\cite{Maxon1967}, with also Elwert correction \cite{Elwert1939} for low temperatures due to the Coulomb effect. The fully relativistic e-e radiation is provided by Haug \cite{Haug1975a}, with also a good approximation for Maxwellian plasma \cite{Haug1975b}. The calculation by Nozawa et al. \cite{Nozawa2009} for e-e radiation has considered all these major effects. After repeating those calculations in the literature, we found that the results provided by van Hoof \cite{Hoof2015} and Nozawa \cite{Nozawa2009} are sufficient as starting points for our present usage, i.e., providing accurate analytical fitting of Bremsstrahlung radiation power for e-i and e-e in fusion plasma applications, with accuracy to within 1\%. We also noticed a recent work by Munirov et al. \cite{Munirov2023}, which has provided comprehensive references on the topic of Bremsstrahlung radiation in plasmas and could be useful for those who are interested. They use relativistic e-i and e-e models to study the suppression of Bremsstrahlung losses from relativistic plasma with energy cutoff, which are not accurate for low energy ranges, and the e-e model is also only an approximation from Ref. \cite{Haug1975b}. Improving the accuracy of Bremsstrahlung calculation is important for advanced fuel fusion, especially for p-B fusion, where a 20\% improvement of the fusion reactivity can significantly change the feasibility of energy gain \cite{Putvinski2019,Xie2023}, which is due to the close relationship between fusion power and Bremsstrahlung loss power.

In the following sections, we revisit the calculations and models in the literature, and choose the models used for the present work in Sec.~\ref{sec:model}. Sec.~\ref{sec:fitting} provides our new analytical fitting for thermal averaged electron temperature. Sec.~\ref{sec:fitenergy} provides analytical fitting for electron energy. Sec.~\ref{sec:summ} summarizes our results.

\section{Revisiting the calculation models}\label{sec:model}

Due to the complicated relativistic and quantum effects, the accurate calculation of Bremsstrahlung radiation is challenging. Many models have been developed. We review some of the most commonly used models in this section. We are mainly interested in the total power loss, and thus the energy spectrum of the radiation can be integrated out. We use SI units here.

Firstly, we summarize some useful notations. The Thomson cross section is given by
\begin{equation}
\sigma_t = \frac{8\pi}{3}r_e^2 = 6.652 \times 10^{-29} \, \rm{m}^2,
\end{equation}
where the classical electron radius is
\begin{equation}
r_e = \frac{e^2}{4\pi\epsilon_0 m_e c^2} = \alpha \frac{hc}{2\pi m_e c^2} = 2.8179 \times 10^{-15} \, \rm{m},
\end{equation}
and the fine structure constant is
\begin{equation}
\alpha = \frac{e^2}{2\epsilon_0 hc} \approx \frac{1}{137.0}.
\end{equation}
Here, $c = 2.99792458 \times 10^8 \, \rm{m/s}$ is the speed of light, $h = 6.626 \times 10^{-34} \, \rm{J}\cdot\rm{s}$ is Planck's constant, $e = 1.602 \times 10^{-19} \, \rm{C}$ is the electron charge, $m_e = 9.1094 \times 10^{-31} \, \rm{kg}$ is the electron mass, $\epsilon_0=8.8542\times 10^{-12} \, \rm{F/m}$ is the vacuum dielectric constant, and $k_B = 1.3806 \times 10^{-23} \, \rm{J/K}$ is the Boltzmann constant.

The Rydberg energy is given by
\begin{equation}
R_y = \frac{m_e e^4}{8\epsilon_0^2 h^2} = \frac{1}{2}\alpha^2 m_e c^2 = 2.17987 \times 10^{-18} \, \rm{J} = 13.6 \, \rm{eV}.
\end{equation}

After integrating over the angles of the outgoing electron and photon, unit time, and unit volume, the radiation photon number between { the energy/frequency range $[\nu,\nu+d\nu]$} is given by \cite{Gould1980}
\begin{eqnarray}\label{eq:Rrate}\nonumber
R &=& \frac{n_1 n_2}{1 + \delta_{12}} \int d^3 \mathbf{p}_1 d^3 \mathbf{p}_2 f_1(\mathbf{p}_1) f_2(\mathbf{p}_2) \cdot\\
&& \left[ (\mathbf{v}_1 - \mathbf{v}_2)^2 - \frac{\mathbf{v}_1 \times \mathbf{v}_2}{c^2} \right]^{1/2} \frac{d\sigma}{d\nu},
\end{eqnarray}
where {$n_{1,2}$, $\mathbf{v}_{1,2}$ and  $\mathbf{p}_{1,2}$ are the number density, velocity and momentum of the two particles respectively, and $\sigma$ is the cross-section}. The distribution function is normalized, i.e., $\int d^3 \mathbf{p}_j f_j(\mathbf{p}_j) = 1$, $j=1,2$, and the Kronecker $\delta_{12}$ is to avoid counting the same collision twice when the collision is between the same type of particles. For e-i collisions, where the ion velocity is much less than the electron velocity, the integral can be simplified by setting ${\bm v}_2=0$. In the non-relativistic case, the integral only depends on the relative velocity $|{\bm v}_1 - {\bm v}_2|$.

\begin{figure*}
\centering
\includegraphics[width=17.5cm]{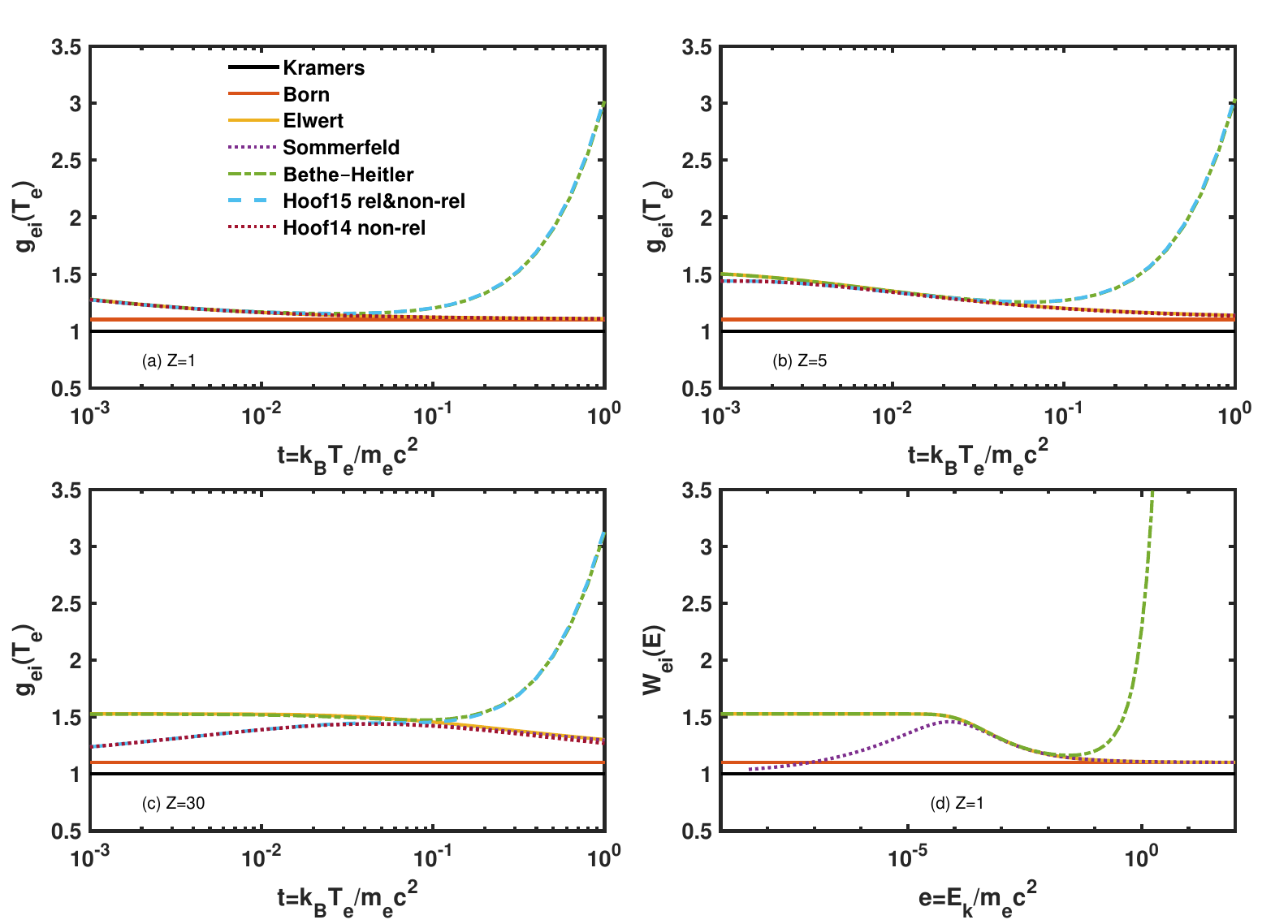}
\caption{Comparison of different models for the thermal average Gaunt factor $g_{ei}$: (a)-(c) show the comparison between various models. Our calculations agree with Hoof's\cite{Hoof2014,Hoof2015} for both non-relativistic and relativistic cases. The values from Hoof15 for both non-relativistic and relativistic cases are considered accurate. The weight factor $W(E)$ for energy is also shown in (d). { It can be observed that the Gaunt factor increases rapidly for $t>0.1$, a behavior attributed to both the relativistic effects and the relativistic distribution function.}}\label{fig:cmpgeimodel}
\end{figure*}

\subsection{Non-relativistic e-i}

The non-relativistic e-i case has been well solved \cite{Berger1957,Greene1959,Karzas1961}. The classical model is due to Kramers, and more accurate ones include the Born approximation and Elwert's correction for low energy. The most accurate one is due to Sommerfeld. The Kramers cross-section is given by\cite{Kramers1923}
\begin{equation}
\frac{d\sigma_k}{d\nu} = \frac{A}{\nu \sqrt{3} \pi} \eta_0^2.
\end{equation}
The Born approximation is expressed as
\begin{equation}
\frac{d\sigma_b}{d\nu} = \frac{A}{\nu \pi^2} \eta_0^2 \ln\left(\frac{\eta_f+\eta_0}{\eta_f-\eta_0}\right).
\end{equation}
The Elwert correction to the Born approximation is given by
\begin{equation}
\frac{d\sigma_e}{d\nu} = \frac{A}{\nu \pi^2} \eta_0 \eta_f \frac{1-e^{-2\pi\eta_0}}{1-e^{-2\pi\eta_f}} \ln\left(\frac{\eta_f+\eta_0}{\eta_f-\eta_0}\right).
\end{equation}
The most accurate Sommerfeld non-relativistic expression is \cite{Sommerfeld1939}
\begin{equation}
\frac{d\sigma_s}{d\nu} = \frac{A}{\nu} \frac{\eta_0^2}{(e^{2\pi\eta_0}-1)(1-e^{-2\pi\eta_f})} x \frac{d}{dx} |F(x)|^2,
\end{equation}
where
\begin{eqnarray}
\eta_{0,f} &=& \left(\frac{Z^2 R_y}{E_{0,f}}\right)^{1/2},~~ x = \frac{4\eta_0 \eta_f}{(\eta_0-\eta_f)^2},\\
v_{0,f} &=& \sqrt{\frac{2E_{0,f}}{m_e}},~~ E_f = E_0 - h\nu,
\end{eqnarray}
and $F(x)$ is the hypergeometric function defined as
\begin{equation}
F(x) = \,_2F_1(i\eta_0, i\eta_f; 1; -x),
\end{equation}
and
\begin{equation}
A = \frac{2\pi\sigma_t}{\alpha} = \frac{2}{3} \frac{he^2}{\epsilon_0 m_e^2 c^3} = 5.728 \times 10^{-26} \, \mathrm{m^2}.
\end{equation}
Note that in some references, the unit $\epsilon_0$ is taken as $\frac{1}{4\pi}$. The radiation photon energy $h\nu \in [0,E_0]$. Here, $E_0$ is the electron energy, and $E_f$ is the outgoing electron energy.

To avoid differentiation, another equivalent equation is given by \cite{Brussaard1962}
{\small\begin{eqnarray}
&&x\frac{d}{dx}|F(x)|^2\\\nonumber
&=&\eta_0\eta_f\frac{|F^2(-i\eta_0+1,-i\eta_f,1;-x)-F^2(-i\eta_f+1,-i\eta_0,1;-x)|}{\eta_f-\eta_0}.
\end{eqnarray}}

We are interested in the radiation power. For a normalized distribution function
\begin{eqnarray}
\int_0^{\infty}f(E_0)dE_0=1,
\end{eqnarray}
the total power spectrum integral is given by
\begin{eqnarray}
P_\nu=n_en_i\int_{h\nu}^{\infty}\Big(\frac{2E_0}{m_e}\Big)^{1/2}h\nu f(E_0)\frac{d\sigma}{d\nu}dE_0.
\end{eqnarray}
For Maxwellian distribution, we have
\begin{eqnarray}
f(E)=\frac{{2}}{\sqrt{\pi}}\frac{1}{(k_BT)^{3/2}}\sqrt{E}e^{-\frac{E}{k_BT}},~~\int_0^{\infty}f(E)dE=1.
\end{eqnarray}

{ Further integrating with respect to all the frequency spectrum yields the total power}
{\small
\begin{eqnarray}\label{eq:RPei}
P{=\int_0^{\infty}P_\nu d\nu}=n_en_i\int_0^{\infty}\int_{h\nu}^{\infty}\Big(\frac{2E_0}{m_e}\Big)^{1/2}h\nu f(E_0)\frac{d\sigma}{d\nu}dE_0d\nu,
\end{eqnarray}
}
where $\Big(\frac{2E_0}{m_e}\Big)^{1/2}$ represents the flux to density conversion factor, and $h\nu$ is the energy of each photon. Equation (\ref{eq:RPei}) is consistent with Equation (\ref{eq:Rrate}).

\subsubsection{For thermal Maxwellian distributions}

It is readily obtained that the Kramers spectrum is given by
\begin{eqnarray}
P_{\nu,k}=n_en_i\frac{32\pi Z^2e^6}{3(4\pi\epsilon_0)^3m_ec^3}\Big(\frac{2\pi}{3m_ek_BT_e}\Big)^{1/2}e^{-\frac{h\nu}{k_BT_e}},
\end{eqnarray}
and the total power is
\begin{eqnarray}\label{eq:Pkrammer}
P=n_en_i\frac{32\pi Z^2e^6}{3(4\pi\epsilon_0)^3hm_ec^3}\Big(\frac{2\pi k_BT_e}{3m_e}\Big)^{1/2}.
\end{eqnarray}
The Born spectrum is given by
\begin{eqnarray}\nonumber
P_{\nu,b}&=&n_en_i\frac{32\pi Z^2e^6}{3(4\pi\epsilon_0)^3hm_ec^3}\Big(\frac{2\pi}{m_ek_BT_e}\Big)^{1/2}\cdot\\
&&e^{-\frac{h\nu}{k_BT_e}}K_0\Big(\frac{h\nu}{2k_BT_e}\Big),
\end{eqnarray}
{ with $K_0$ be the modified Bessel function of the second kind,} and the total power is
\begin{eqnarray}
P=n_en_i\frac{64Z^2e^6}{3(4\pi\epsilon_0)^3m_ec^3}\Big(\frac{2\pi k_BT_e}{m_e}\Big)^{1/2},
\end{eqnarray}
which is approximately $\frac{2\sqrt{3}}{\pi}\simeq1.10$ times larger than Kramers' result in Eq.(\ref{eq:Pkrammer}). Here, we have used
\begin{eqnarray}
\int_0^{\infty}e^{-ax}\ln(\sqrt{x+1}+\sqrt{x})dx=\frac{1}{2a}e^{a/2}K_0\Big(\frac{a}{2}\Big),
\end{eqnarray}
and
\begin{eqnarray}
\int_0^{\infty}e^{-ax}K_0(ax)dx=\frac{1}{a}.
\end{eqnarray}
For the Elwert correction Born approximation and Sommerfeld model, the integral can only be calculated numerically. The above results agree with Refs.\cite{Bekefi1966,Berger1957,Greene1959,Karzas1961}.

{It is noteworthy that while the total power may vary only slightly with different Gaunt factors, the spectrum can show significant differences among various models. Therefore, for those interested in an accurate spectrum, it is essential to utilize the most precise model available.}

\subsubsection{For arbitrary distributions}

For arbitrary distributions, the total power can be obtained by first integrating over frequency and then integrating over energy, i.e.,
\begin{eqnarray}
P=n_en_i\int_{0}^{\infty}\int_0^{E_0}\Big(\frac{2E_0}{m_e}\Big)^{1/2}h\nu f(E_0)\frac{d\sigma}{d\nu}d\nu dE_0.
\end{eqnarray}
That is
\begin{eqnarray}
Q_E=\int_0^{E_0}h\nu \frac{d\sigma}{d\nu}d\nu,\\
P=n_en_i\int_{0}^{\infty}\Big(\frac{2E_0}{m_e}\Big)^{1/2}Q_E f(E_0)dE_0.
\end{eqnarray}
Define
\begin{eqnarray}
w=\frac{\nu\sqrt{3}\pi}{AR_yZ^2}\frac{d\sigma}{d\nu},
\end{eqnarray}
which gives
\begin{eqnarray}
w_k=\frac{1}{E_0},\\
w_b=\frac{\sqrt{3}}{\pi}\frac{1}{E_0}\ln\Big(\frac{\sqrt{E_0}+\sqrt{E_f}}{\sqrt{E_0}-\sqrt{E_f}}\Big).
\end{eqnarray}
And define the frequency integral weight
\begin{eqnarray}
W_E(E_0)=\int_0^{E_0}wdh\nu,
\end{eqnarray}
which yields,
\begin{eqnarray}
W_{E,k}(E_0)=1,\\
W_{E,b}(E_0)=\frac{2\sqrt{3}}{\pi},
\end{eqnarray}
where we have used $\int_0^a[2\ln(\sqrt{a}+\sqrt{a-x})-\ln x]dx=2a$. The Elwert $W_{E,e}(E_0)$ and Sommerfeld $W_{E,s}(E_0)$ can only be calculated numerically.

For arbitrary distributions, the total power is 
\begin{eqnarray}\nonumber
P&=&n_en_i\frac{AR_yZ^2}{\sqrt{3}\pi}\int_{0}^{\infty}\Big(\frac{2E_0}{m_e}\Big)^{1/2}W_E(E_0)f(E_0)dE_0\\
&=&n_en_i\frac{16\pi^2Z^2e^6}{3\sqrt{3}(4\pi\epsilon_0)^3hm_ec^3}\langle v_0\rangle_W,
\end{eqnarray}
where
\begin{eqnarray}
\langle v_0\rangle_W\equiv\int_0^{\infty}\Big(\frac{2E_0}{m_e}\Big)^{1/2}W_E(E_0)f(E_0)dE_0,
\end{eqnarray}
is the weight averaged velocity. For Kramers and Born model, the weight $W(E_0)$ is constant, and thus the radiation power is proportional to the average velocity. For Maxwellian plasma, the averaged velocity is
\begin{eqnarray}
\langle v_0\rangle=\Big(\frac{8k_BT_e}{\pi m_e}\Big)^{1/2}.
\end{eqnarray}

For Maxwellian plasma, define Gaunt factor\cite{Gaunt1930}
\begin{eqnarray}
g_{ei}=\frac{\langle v_0\rangle_W}{\langle v_0\rangle}=\Big(\frac{8k_BT_e}{\pi m_e}\Big)^{-1/2}\langle v_0\rangle_W,
\end{eqnarray}
which gives
\begin{eqnarray}
P=n_en_i\frac{32\pi Z^2e^6}{3(4\pi\epsilon_0)^3hm_ec^3}\Big(\frac{2\pi k_BT_e}{3m_e}\Big)^{1/2}g_{ei}.
\end{eqnarray}
For quasi-neutral plasma $n_e=\sum_i Z_i n_i$. Hence we have roughly $P\propto Zn_e^2$, i.e., ions with higher atomic numbers $Z$ yield stronger radiation.

We numerically calculated Sommerfeld's model, and the results agree well with van Hoof\cite{Hoof2014}. Note also that the non-relativistic $g_{ei}(t,Z)=g_{ei}(t/Z^2)$, i.e., the effect of $Z$ can be absorbed into the temperature $t=\frac{k_BT_e}{m_ec^2}$. Also note that $g_{ei}^{nr}(t\to0)=1$ and $g_{ei}^{nr}(t\to\infty)=\frac{2\sqrt{3}}{\pi}$, with a maximum at around 1.46 (the Elwert Born is around 1.53).

\begin{figure*}
\centering
\includegraphics[width=17.5cm]{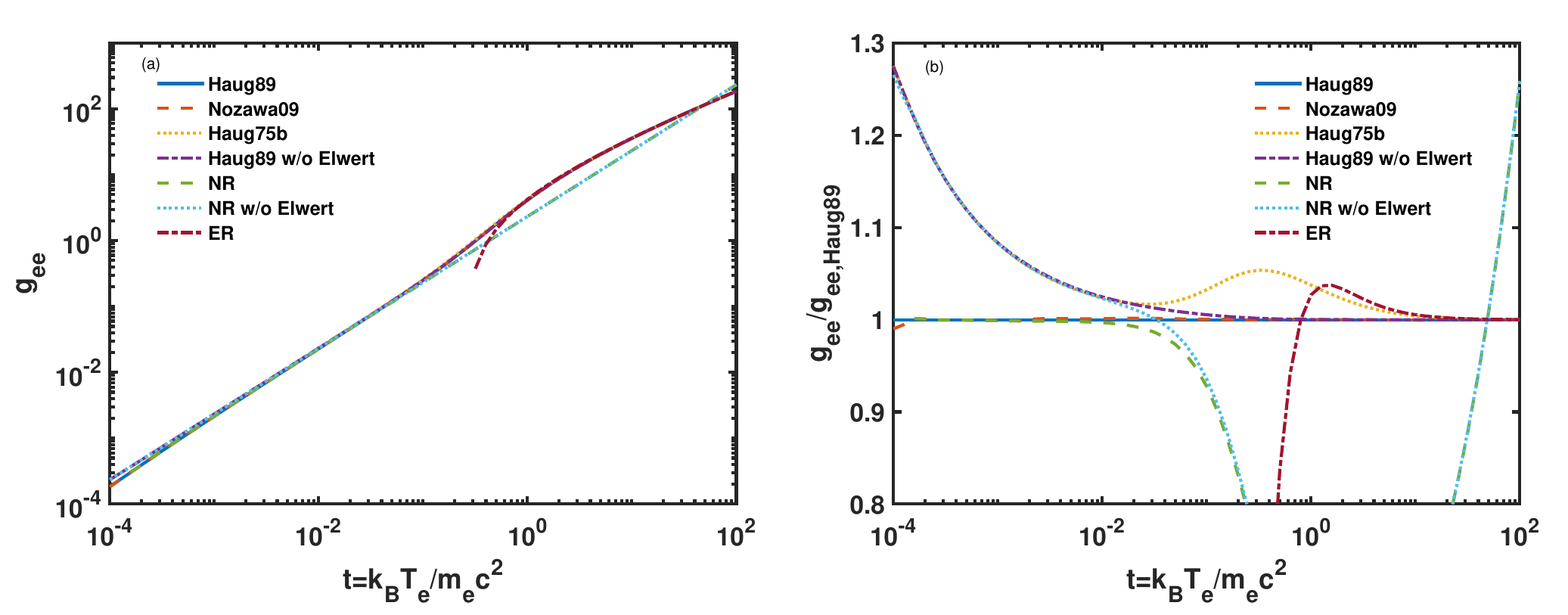}
\caption{Comparison of different models for the thermal average Gaunt factor $g_{ee}$. Our calculations based on Haug75b\cite{Haug1975b} and Haug89\cite{Haug1989} are consistent with Nozawa's\cite{Nozawa2009} and can be reduced to the non-relativistic (NR) and extreme-relativistic (ER) cases. Haug75b shows an error of around 5\% at 100 keV (b). { Therefore, for those requiring highly accurate calculations, the commonly used results from Haug75b should be approached with caution.}}\label{fig:cmpgeemodel}
\end{figure*}

\begin{figure*}
\centering
\includegraphics[width=17.5cm]{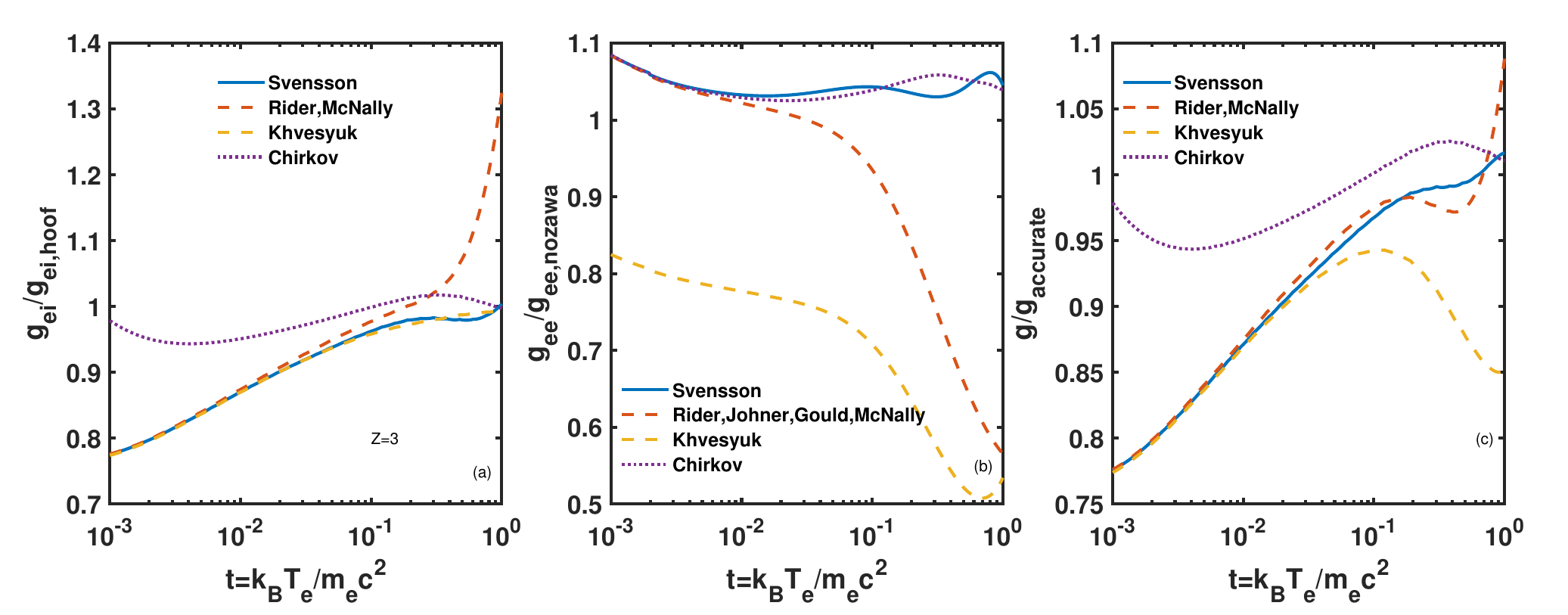}
\caption{Comparison of thermal average fittings in the literature for (a) e-i with the accurate data from Hoof\cite{Hoof2015} (for $Z=3$), and (b) e-e with the accurate data\cite{Nozawa2009,Haug1989}. The errors are commonly larger than 5\%. The total $g$ is also shown (c). { These results suggest that previous fittings should be used with caution for those seeking highly accurate calculations.}}\label{fig:cmpfitting_Z=3}
\end{figure*}

\begin{figure*}
\centering
\includegraphics[width=17.5cm]{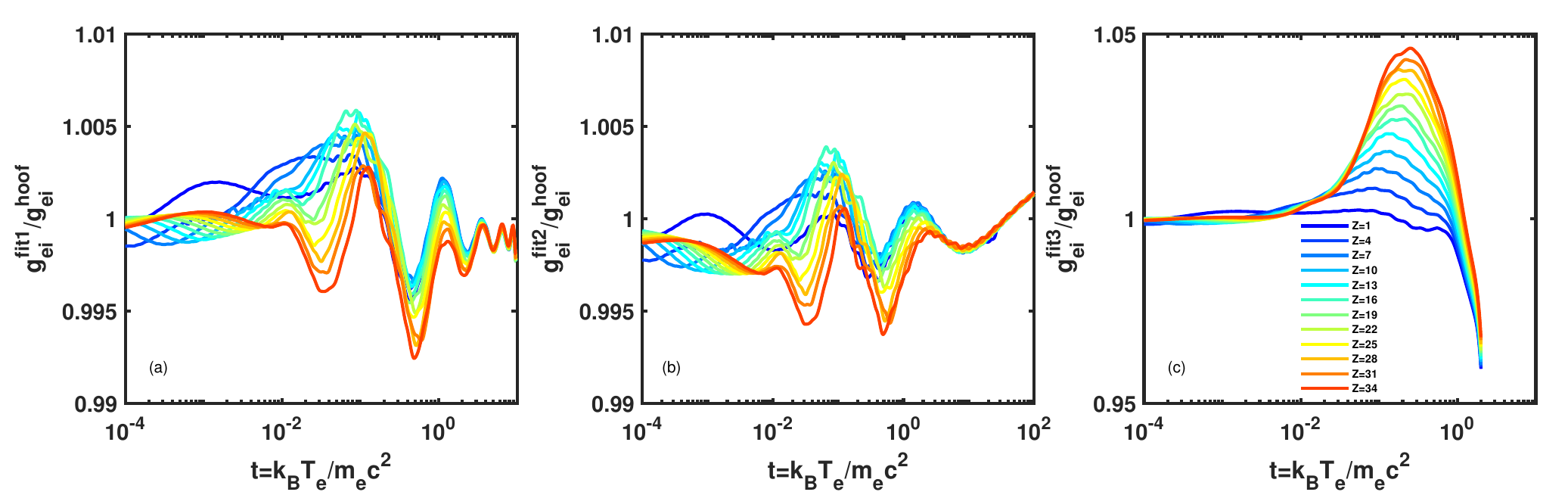}
\caption{Comparison of the three new fittings of thermal averaged $g_{ei}$ with Hoof15, showing (a) fitting \#1 Eq.(\ref{eq:geifit1}) with an error less than 1\%, for $t \leq 10$ and $Z \leq 34$, (b) fitting \#2 Eq.(\ref{eq:geifit2}) with an error less than 1\%, for all $t$, and (c) fitting \#3 Eq.(\ref{eq:geifit3}) with an error less than 5\%, for $t \leq 2$.  { These new fittings have carefully addressed the effects of $Z$, the behavior as $t\to0$, and relativistic effects, making them more accurate than previous fittings across a wide range of conditions.}}\label{fig:geifit_xie}
\end{figure*}

\begin{figure*}
\centering
\includegraphics[width=17.5cm]{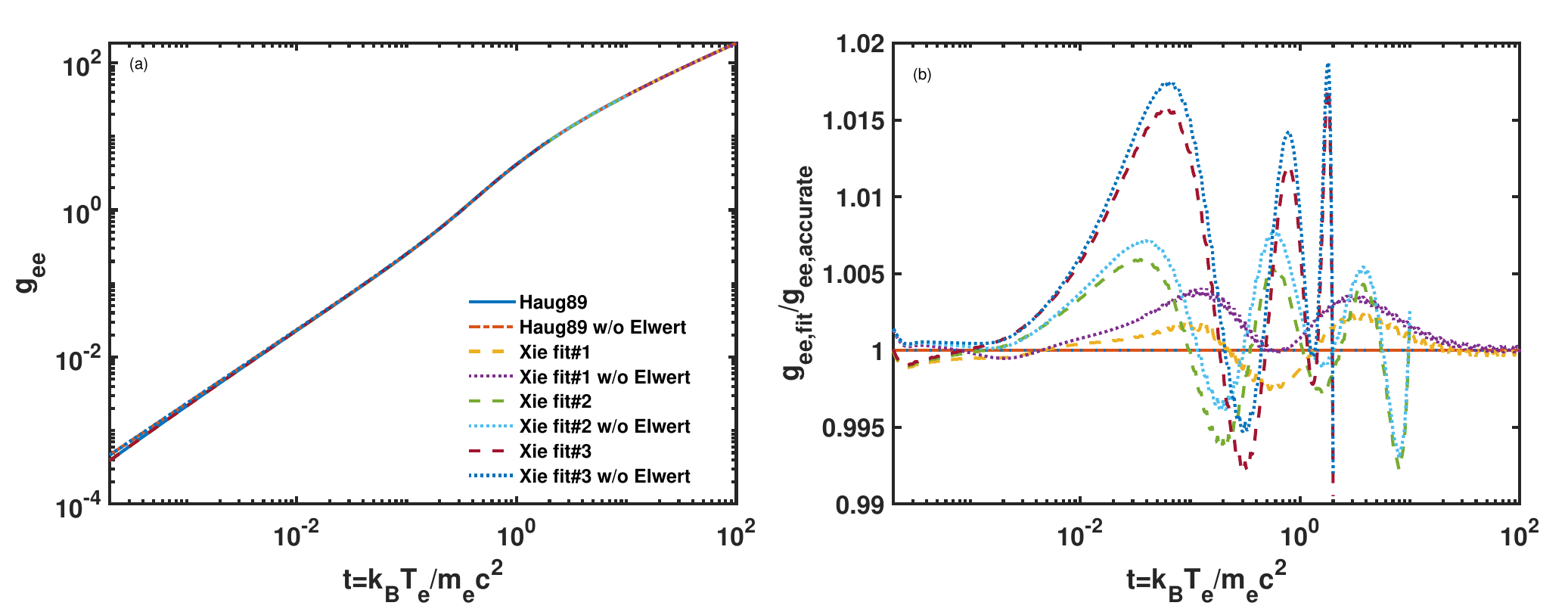}
\caption{Comparison of the new fittings of thermal averaged $g_{ee}$ with accurate data\cite{Nozawa2009,Haug1989}. Both with and without Elwert factor are fitted with errors of $\lesssim$1\%, 2\%, and 0.5\% for Eqs. (\ref{eq:geefit1}) [\#1], (\ref{eq:geefit2}) [\#2], and (\ref{eq:geefit3}) [\#3], respectively, for $T_e \geq 0.1$ keV. { These new fittings have carefully addressed both the small $t\to0$ and intermediate $t\simeq1$ regions, resulting in improved accuracy compared to previous fittings across a broad range of conditions.}}\label{fig:geefit_xie}
\end{figure*}

\subsection{Relativistic e-i}\label{sec:geirel}

The usual relativistic e-i model used is from Bethe–Heitler (BH) with Born approximation, which is valid for high energy but not correct for low electron energy and low photon energies because the Coulomb distortion of the wave functions becomes too large, and the Born approximation breaks down. Using the notation of Chluba et al\cite{Chluba2020}, the classical Kramers model is given by
\begin{eqnarray}
\frac{d\sigma_k(\omega,p_0)}{d\omega}=\frac{2\alpha Z^2}{\sqrt{3}}\frac{\sigma_t}{p_0^2\omega}=\frac{16\pi\alpha Z^2}{3\sqrt{3}}\frac{r_e^2}{p_0^2\omega}.
\end{eqnarray}
The first-order relativistic Born approximation model (BH)\cite{Bethe1961} with Elwert correction is
\begin{eqnarray}
\frac{d\sigma_{BH}}{d\omega} = \frac{d\sigma_{k}(\omega,p_0)}{d\omega}\frac{\eta_{f}}{\eta_0}\frac{(1 - e^{-2\pi\eta_0})}{(1 - e^{-2\pi\eta_{f}})}g_{BH}(\omega,p_0),\\\nonumber
g_{BH}(\omega,p_0) = \frac{\sqrt{3}}{\pi}\Big[\frac{p_0p_{f}}{4} - \frac{3}{8}\gamma_0\gamma_f\Big(\frac{p_0}{p_{f}}+\frac{p_{f}}{p_0}\Big) + \gamma_0\gamma_fL+\\\nonumber
\frac{3}{8}\omega L\Big[\Big(1+\frac{\gamma_0\gamma_f}{p_0^2}\Big)\frac{\lambda_0}{p_0}-\Big(1+\frac{\gamma_0\gamma_f}{p_f^2}\Big)\frac{\lambda_{f}}{p_{f}}+
\omega\Big(1+\frac{\gamma_0\gamma_f}{p_0^2p_{f}^2}+\\
\frac{\gamma_0^2\gamma_f^2}{p_0^2p_{f}^2}\Big)\Big]+
\frac{3}{8}\Big(\frac{\gamma_fp_{f}}{p_0^2}\lambda_0+\frac{\gamma_0p_0}{p_{f}^2}\lambda_{f}-2\lambda_0\lambda_{f}\Big)\Big], 
\end{eqnarray}
where
\begin{eqnarray*}
\lambda_i = \ln(\gamma_i + p_i), ~~ L = \ln\Big(\frac{\gamma_0\gamma_f + p_0p_{f} - 1}{\omega}\Big), \\
\omega = \frac{\hbar\nu}{m_{e}c^2}, ~~ \eta_i = \frac{aZ\gamma_i}{p_i},~~
\gamma = (1 + p^2)^{1/2}, \\ E = \sqrt{p^2c^2 + (m_{e}c^2)^2} = \gamma m_{e}c^2,~~
p_{f} = \sqrt{p_0^2 + \omega(\omega - 2\gamma_0)}.
\end{eqnarray*}
Here, $\hbar=h/(2\pi)$, $p_i$ and $\gamma_i$ have been normalized with $m_ec$ and $m_ec^2$. Note that the electron velocity is given by $v=pc^2/E$, and the kinetic energy by $E_k=m_ec^2(\gamma-1)$.

Again, we define
\begin{eqnarray}\label{eq:wE}
w=\frac{\omega\sqrt{3}\pi}{ARZ^2}\frac{d\sigma}{d\omega}, ~~W_E(E_{k0})=\int_0^{E_{k0}}wdh\nu,
\end{eqnarray}
which gives
\begin{eqnarray}
w_{BH}=\frac{1}{m_ec^2}\frac{2}{p_0^2}\frac{\eta_{f}}{\eta_0}\frac{(1 - e^{-2\pi\eta_0})}{(1 - e^{-2\pi\eta_{f}})}g_{BH}(\omega,p_0).
\end{eqnarray}

For arbitrary distribution, the total power is
\begin{eqnarray}\nonumber
P&=&n_en_i\frac{AR_yZ^2}{\sqrt{3}\pi}\int_{0}^{\infty}p_0^2W_E|E_{rel}|f(p_0)dp_0\\\nonumber
&=&n_en_i\frac{AR_yZ^2}{\sqrt{3}\pi}\int_{0}^{\infty}W_E\frac{cp_0^3}{\gamma_0}f(p_0)dp_0\\
&=&n_en_i\frac{16\pi^2Z^2e^6}{3\sqrt{3}(4\pi\epsilon_0)^3hm_ec^3}\langle p_0c\rangle_W,
\end{eqnarray}
where
\begin{eqnarray}
|E_{rel}|=\frac{cp_0}{\gamma_0},~~\int_0^{\infty}p_0^2f(p_0)dp_0=1,
\end{eqnarray}
and 
\begin{eqnarray}
\langle p_0c\rangle_W=\int_{0}^{\infty}W_E\frac{cp_0^3}{\gamma_0}f(p_0)dp_0,
\end{eqnarray}
is the weight averaged velocity.

The relativistic Maxwellian distribution is given by
\begin{eqnarray}
f(p)=\frac{e^{-\gamma(p)/t}}{\theta K_2(1/t)},~~\gamma=(1+p^2)^{1/2}, ~~t=k_BT/mc^2,
\end{eqnarray}
where $K_2$ is the modified Bessel function of the second kind. Reduced to the non-relativistic case, we have
\begin{eqnarray}
f(p)\simeq\frac{\sqrt{2}e^{-p^2/2t}}{\pi^{1/2}t^{3/2}},
\end{eqnarray}
where $K_\alpha(z\to\infty)\to\sqrt{\frac{\pi}{2z}}e^{-z}$. Hence the average velocity is
\begin{eqnarray}\nonumber
\langle p_0c\rangle&=&\int_{0}^{\infty}\frac{cp_0^3}{\gamma_0}f(p_0)dp_0=c\frac{2t(t+1)e^{-1/t}}{K_2(1/t)}\\
&\simeq& c\Big(\sqrt{\frac{8}{\pi}}\sqrt{t}-\frac{7}{\sqrt{8\pi}}\sqrt{t^3}+\cdots\Big).
\end{eqnarray} 
Define the Gaunt factor as
\begin{eqnarray}
g_{ei}=\frac{\langle p_0\rangle_W}{\langle p_0\rangle_M},~~\langle p_0\rangle_M=\sqrt{\frac{8}{\pi}}\sqrt{t},
\end{eqnarray}
where $\langle p_0\rangle_M$ is the non-relativistic lowest order average velocity normalized by $c$. The relativistic Gaunt factor comes from both the relativistic effect and the distribution function, i.e., the relativistic distribution has changed the relation between temperature and the average velocity. The calculation of the relativistic BH model could be even simpler than the non-relativistic Sommerfeld model, since no complicated special function such as the hypergeometric function is involved. Chluba et al \cite{Chluba2020} use a more complicated five-dimensional integral EH\cite{Elwert1969} model which can reduce to BH and Sommerfeld, showing that the direct combination of these two models as in Hoof\cite{Hoof2015} can still be a good approximation for most regions. Hence, we use the data in Hoof\cite{Hoof2015} as our starting point. Our calculation agrees well with Hoof\cite{Hoof2015}, as shown in Fig.\ref{fig:cmpgeimodel}. The comparison of different models is also shown.

For the extreme relativistic (ER) case ($t\gg1$), the above result reduces to\cite{Maxon1972}
\begin{eqnarray}\label{eq:geiER}
g_{ei}^{ER}(t)=\frac{9\sqrt{6}}{8\sqrt{\pi}}t^{1/2}\Big[\ln(2t)+\frac{3}{2}-C_E\Big],
\end{eqnarray}
where $C_E=0.5772$ is the Euler constant.

\subsection{Non-relativistic e-e}
The e-e radiation can also be calculated via Gaunt factor:
\begin{eqnarray}
g_{ee}(t,u)=t J(t,u).
\end{eqnarray}
For Maxwellian plasma\cite{Maxon1967,Itoh2001}:
\begin{eqnarray}
J(t,u)=\frac{\sqrt{3}}{10\sqrt{2}\pi}u^2e^uI(t,u),~~t=\frac{k_BT_e}{m_ec^2},~~u=\frac{h\nu}{k_BT_e},
\end{eqnarray}
\begin{eqnarray}
I(t,u)=\int_0^1dyF_{ee}A(u,y),
\end{eqnarray}
\begin{eqnarray}\nonumber
A(u,y)&=&\frac{e^{-u/y}}{y^3}\Big\{\Big[17-\frac{3y^2}{(2-y)^2}\Big]\sqrt{1-y}+\ln\Big(\frac{1}{\sqrt{y}}+\sqrt{\frac{1}{y}-1}\Big)\\
&&\cdot\Big[\frac{12(2-y)^4-7(2-y)^2y^2-3y^4}{(2-y)^3}\Big]\Big\},
\end{eqnarray}
\begin{eqnarray}
F_{ee}(u,y)=\frac{1}{\sqrt{1-y}}\frac{\exp\Big(\pi\alpha\sqrt{\frac{y}{t u}}\Big)-1}{\exp\Big(\pi\alpha\sqrt{\frac{y}{t u (1-y)}}\Big)-1},
\end{eqnarray}
where the integral term $F_{ee}$ in $I(t,u)$ comes from Elwert's Coulomb correction for low energy\cite{Maxon1967}, which is usually omitted\cite{Maxon1972,Haug1975b} for simplicity.

After integrating over frequency:
\begin{eqnarray}
g_{ee}(t)=t \int_0^{\infty}e^{-u}J(t,u)du=\frac{\sqrt{3}}{10\sqrt{2}\pi}\int_0^{\infty}u^2I(t,u)du.
\end{eqnarray}
If we omit the Elwert correction, i.e., set $F_{ee}=1$, the Gaunt factor can be integrated out analytically to\cite{Maxon1967,Haug1975b,Stickforth1961}:
\begin{eqnarray}
g_{ee}(t)=g_0^{NR}=\frac{2\sqrt{3}}{\pi}\frac{5(44-3\pi^2)}{24\sqrt{2}}t\simeq\frac{2\sqrt{3}}{\pi}\frac{3}{\sqrt{2}}t,
\end{eqnarray}
where $5(44-3\pi^2)/9\simeq7.995\simeq8$ is used.

\subsection{Relativistic e-e}
The comprehensive relativistic e-e calculation\cite{Haug1975a} contains a five-dimensional integral with complicated terms. A reduced integral, still accurate, is provided by Haug\cite{Haug1989}, which we summarized in \ref{sec:dsgmdkee} and will use it later. Here we list Haug's\cite{Haug1975b,Johner1987} simplified two-dimensional integral calculation for Maxwellian plasmas with an error less than 6\% for reference. The Gaunt factor is:
\begin{eqnarray}\nonumber
g_{ee}(t)&=&g_0^{NR}\frac{9\sqrt{\pi}}{20(44-3\pi^2)}\frac{1}{t^{7/2}\Big[e^{1/t}K_2(1/t)\Big]^2}
\int_0^{\infty}du_1\cdot\\
&&e^{-u_1/t}\int_0^{\infty}(u_1+u_2+2)e^{-u_2/t}J(u_1,u_2)du_2,
\end{eqnarray}
and
\begin{eqnarray*}
J(u_1,u_2)=I(w_1w_2+p_1p_2)-I(w_1w_2-p_1p_2),
\end{eqnarray*}
\begin{eqnarray*}
I(\mu)&=&(\mu/2-2)\sqrt{\mu^2-1}-\frac{11}{12}\mu^2+\frac{20}{3}\mu-\frac{8}{3}\ln(\mu+1)+
\\&&
\Big[\frac{3}{2}+\Big(\frac{\mu}{2}-\frac{8}{3}\frac{\mu+2}{\mu+1}{\color{red}\Big)}\sqrt{\mu^2-1}\Big]\ln(\mu+\sqrt{\mu^2-1})+
\\&&
\frac{7}{4}[\ln(\mu+\sqrt{\mu^2-1})]^2,
\end{eqnarray*}
\begin{eqnarray*}
u_i=w_i-1,~~p_i=\sqrt{w_i^2-1},~~t=\frac{k_BT_e}{m_ec^2},
\end{eqnarray*}
where $p_i$ is the momentum normalized by $m_ec$, and $g_0^{NR}$ is the non-relativistic value. For the weak relativistic case, the above result reduces to $g_0^{NR}$. For the extreme relativistic case, the above result reduces to\cite{Maxon1972,Haug1975b,Alexanian1968}:
\begin{eqnarray}\label{eq:geeER}
g_{ee}^{ER}(t)=\frac{9\sqrt{6}}{4\sqrt{\pi}} t^{1/2}\Big[\ln(2t)+\frac{5}{4}-C_E\Big],
\end{eqnarray}
which is around twice of $g_{ei}^{ER}$ for $Z=1$. We solved both this simplified approximated model\cite{Haug1975b} and the complicated accurate model\cite{Haug1989}.

Nozawa et al \cite{Nozawa2009} [note: their $G(t)\cdot t=\sqrt{\frac{8}{3\pi}}g_{ee}$] have solved the comprehensive relativistic e-e calculation by Haug\cite{Haug1975a}, and also used the Elwert correction calculated in Ref.\cite{Itoh2001,Itoh2002} for $T_e<1$keV. We compare our calculations using the above non-relativistic\cite{Maxon1967,Itoh2001} and relativistic\cite{Haug1975b,Haug1989} models in Fig.\ref{fig:cmpgeemodel}, and show consistent results with \cite{Nozawa2009}. Hence, we trust our calculations and the results in \cite{Nozawa2009,Haug1989}, and use them as our start point for our fitting. The fitting in \cite{Nozawa2009} is separated into several regions and would be complicated if we only need a simple formula.

\subsection{Accuracy of existing fitting}
After the above benchmark and confirmation of e-e in Nozawa09\cite{Nozawa2009} and e-i in Hoof15\cite{Hoof2015}, we assume them are accurate, say with an error less than 0.5\%, for our standard normalized cases. Fig.\ref{fig:cmpfitting_Z=3} shows the comparison of different fittings in Fig.\ref{fig:existfitting}. Svensson's \cite{Svensson1982} fitting has an overall good fit, however, with an error $>10$\% at $T_e\simeq20$keV for even $Z=3$. Khvesyuk's \cite{Khvesyuk1998} fitting also does not improve too much. Chirkov's \cite{Chirkov2010} fitting is better due to the correction term for low temperature. The lowest-order e-e fitting has an error less than 10\% for $T_e\leq60$keV.

Coincidentally, the proton-Boron plasma with $Z_{\rm{eff}}\simeq3$, Rider's total fitting $g=Zg_{ei}+g_{ee}$ ($g_{ei}$ is overestimated, and $g_{ee}$ is underestimated) is close to the accurate value. Thus, the conclusions in \cite{Cai2022,Xie2023,Liu2024,Nevins1998} do not change too much.

\begin{figure*}
\centering
\includegraphics[width=17.5cm]{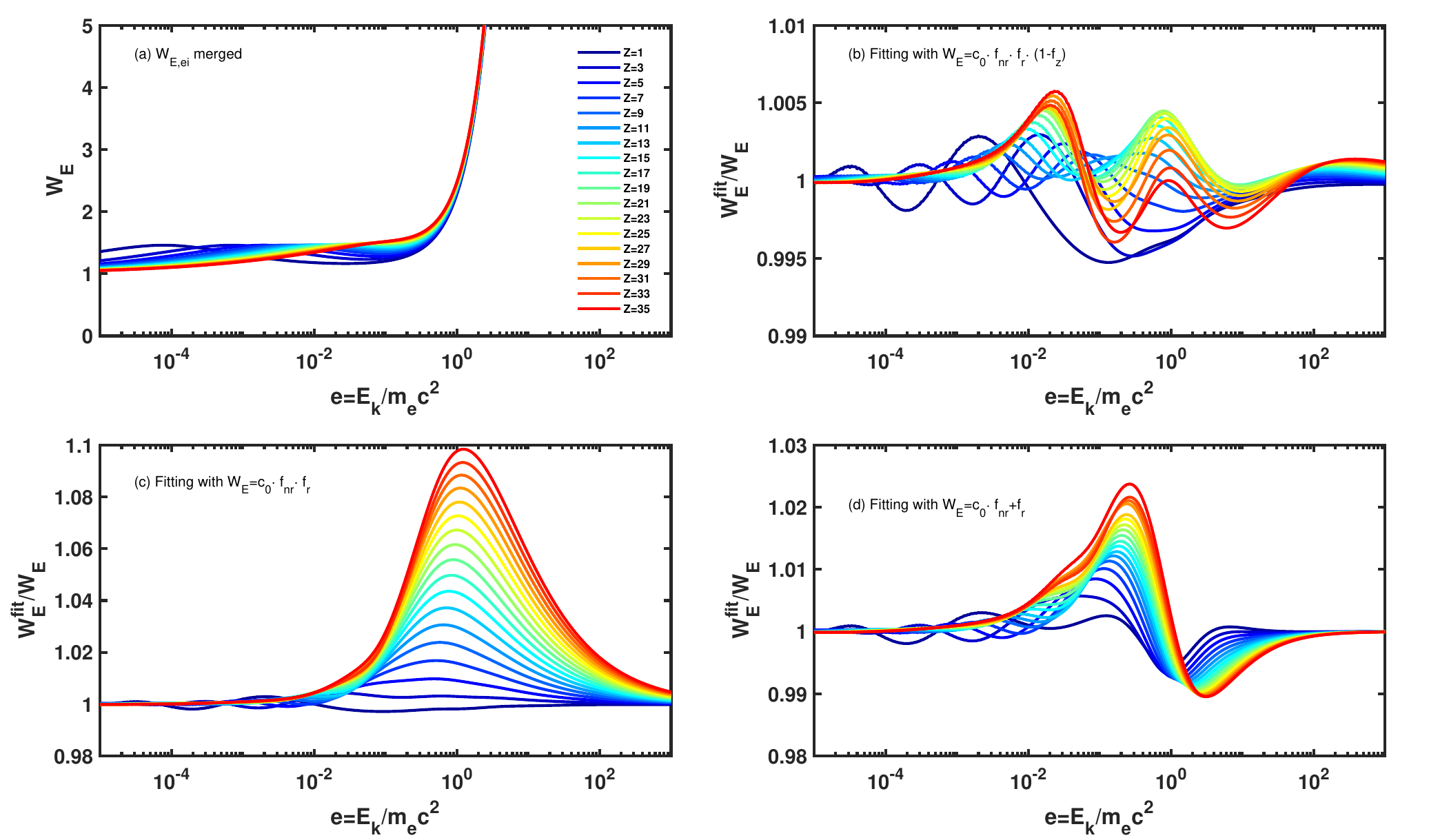}
\caption{Comparison of the new fitting of $W_{ei}(E)$ with accurate merged data, with errors less than 1\%, for $Z\leq35$. The fitting without $f_z$ is also shown in (c). { These new fittings have carefully addressed the effects of $Z$, as well as the conditions for $t\to0$ and relativistic effects, making them accurate across a wide range of conditions.}}\label{fig:WEeifit_xie}
\end{figure*}

\begin{figure*}
\centering
\includegraphics[width=17.5cm]{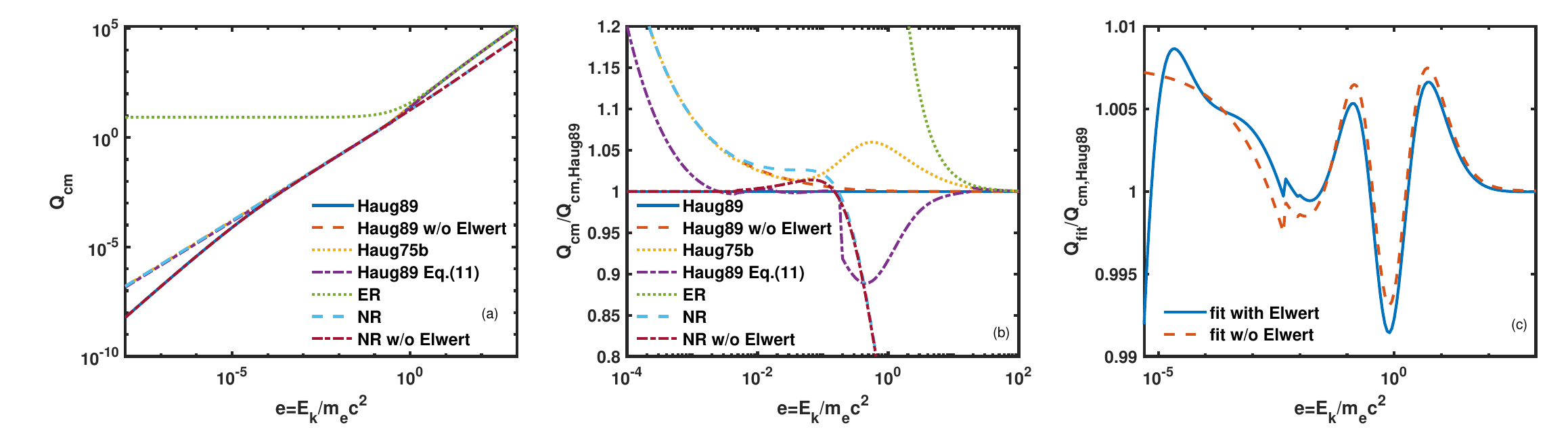}
\caption{Comparison of the $Q_{cm}$ for e-e in different models, with Haug89 being the most accurate. The new fitting given by Eq.(\ref{eq:Qcmfit}) achieves errors less than 1\%. { These new fittings have carefully addressed both the small $t\to0$ and intermediate $t\simeq1$ regimes, resulting in improved accuracy compared to previous fittings across a wide range of conditions.}}\label{fig:cmpQcm}
\end{figure*}

\section{Analytical Fitting for thermal averaged electron temperature}\label{sec:fitting}

After the knowledge learned from Section \ref{sec:model}, we can fit firstly the accurate non-relativistic part and extreme-relativistic part, and then fit the middle range, which can keep all regions accurate. The total radiation power for Maxwellian quasi-neutral plasma is
\begin{eqnarray}\label{eq:ptotal}
P=\frac{32\pi e^6n_e^2}{3(4\pi\epsilon_0)^3hm_ec^3}\Big(\frac{2\pi k_BT_e}{3m_e}\Big)^{1/2}[Zg_{ei}(t,Z)+g_{ee}(t)].
\end{eqnarray}

In this section, we provide the fittings with truncated polynomial to $t\leq10$ to error less than 1\%. Some more fittings will be listed in \ref{sec:morefit}.

\subsection{Fitting for e-i}

We firstly fit the non-relativistic term $f_{nr}$, which depends on $t/Z^2$. Then we fit the remaining relativistic term $f_{r}$ for $Z=1$. And finally, we add the small term $f_z$ for $Z\neq1$. For $t\leq10$, we have ($g_{ei}^{fit\#1}$)
\begin{eqnarray}\label{eq:geifit1}
g_{ei}(t,Z)=c_0(1+f_{nr}-f_z)+f_{r},
\end{eqnarray}
\begin{eqnarray}\label{eq:geinr}\nonumber
f_{nr}&=&c_{nr1}\Big\{1-\exp\Big[-\Big(c_{nr2}\frac{Z^2}{t}\Big)^{c_{nr4}}\Big]\Big\}-\\
    &&(c_{nr1}+c_{nr6})\exp\Big[-\Big(\frac{t}{c_{nr3}Z^2}\Big)^{c_{nr5}}\Big],
\end{eqnarray}
\begin{eqnarray}\label{eq:geir}
f_{r}=c_0\sum_{j=1}^{10}c_{rj}t^j,
\end{eqnarray}
\begin{eqnarray}
f_z=\frac{\Big(\frac{Z}{10}\Big)c_{z1}\Big(100t\sqrt{\frac{10}{Z}}\Big)^{c_{z2}}}{\exp\Big[c_{z3}\Big(100t\sqrt{\frac{10}{Z}}\Big)^{c_{z4}}\Big]-1}.
\end{eqnarray}
where
\begin{eqnarray*}
c_{nr}&=&[0.4302,~24.2255e-5,~0.7546e-5,~ 0.5282,~\\
&&  0.3301,~ 0.0911],\\
c_{r}&=&[0.55467,~    2.6346,~    -2.277595,~    1.1480,~\\&&    -0.36465,~    0.07451,~    -0.00975,~    0.0007885,\\&&    -3.5841e-5,~    6.99834e-7],\\ 
c_z&=& [5.760e4,~3.440,~16.80,~0.1333].
\end{eqnarray*}
For $t>10$, we use $g_{ei}(t)=g_{ei}^{ER}(t)$ in Eq.(\ref{eq:geiER}) for the fitting.

For low accuracy with error less than 5\%, using only $f_{nr}$ and sixth-order polynomial $f_r$ can be used for $t\leq2$, as given in \ref{sec:fitpoly}. Additionally, a more complicated form to fit all ranges to high accuracy is provided in \ref{sec:fitall}.

The results are shown in Fig.\ref{fig:geifit_xie}, with a total error less than 1\% for $Z=1-35$, as compared with Hoof15. For small $Z\leq5$, the $f_z$ term can be omitted to keep the error still less than 2\%. Even for $Z=36$, the maximum contribution due to $f_z$ is less than 5\%. The $f_{nr}$ term fits the non-relativistic Sommerfeld's model perfectly with an error less than $0.2$\%. A major difference of our fitting to previously used fittings is that we keep $g_{ei}\to1$ as the correct value for $t\to0$, not the incorrect $g_{ei}^{nr}\to {2\sqrt{3}}/{\pi}$. This feature is also treated correctly in Chirkov\cite{Chirkov2010}. The other two fitting forms are also shown in Fig.\ref{fig:geifit_xie} with the error as stated.

\subsection{Fitting for e-e}

It is difficult to use a polynomial alone (less than tenth order) to fit all the effects of Elwert correction at low temperature, non-relativistic factor, and relativistic effect to high accuracy. Hence, we use the combination of three factors to do the fitting for $t\leq10$. We use
\begin{eqnarray}\label{eq:geefit1}
g_{ee}(t)=F_{ee}F_{NR}(1+0.53t+9.48t^2-0.67t^3+0.027t^4),
\end{eqnarray}
with
\begin{eqnarray}\label{eq:Feet}
F_{ee}(t)=\frac{1}{2}\Big\{\tanh\Big[0.602(\log_{10}t+5.06)\Big]+1\Big\},
\end{eqnarray}
and
\begin{eqnarray}
F_{NR}(t)=t\frac{6\sqrt{3}}{\sqrt{2}\pi}\frac{1}{2}\Big\{\tanh\Big[-2.153\log_{10}\Big(\frac{t}{0.43}\Big)\Big]+1\Big\}.
\end{eqnarray}
The above Elwert correction factor $F_{ee}$ is rather accurate to an error less than 0.2\%. The $F_{NR}$ with $\tanh$ function is to separate the non-relativistic term since it is proportional to $t$. The remaining polynomial term is to fit the relativistic effect. 

For low accuracy with an error less than 2\%, using only $F_{ee}$ and a fifth-order polynomial can be used for $t\leq2$, as given in \ref{sec:fitpoly}. Additionally, a more complicated form for fitting all ranges to high accuracy is also provided in \ref{sec:fitall}.

The results are shown in Fig.\ref{fig:geefit_xie}, with an overall error less than 1\% compared with Nozawa09. The Elwert correction is only important at low energy range, say, $T_e\leq10$keV, where the e-e is far smaller than e-i. Thus, high accuracy is not a must and one can also set $F_{ee}=1$ for simplification. For $t>10$, one can use $g_{ee}(t)=g_{ee}^{ER}(t)$ in Eq.(\ref{eq:geeER}) for the fitting.

\section{Analytical Fitting for electron energy}\label{sec:fitenergy}
To study the arbitrary distribution of electrons, the fitting of the Bremsstralung cross-section for energy could be useful to reduce the calculations.

\subsection{Fitting for e-i}
From Eq.(\ref{eq:wE}), we need only fitting $W_{E}(E,Z)$. We use a similar method as in Hoof15\cite{Hoof2015} to merge the non-relativistic result and BH relativistic result and then give the fitting. The merged result is set as 
\begin{eqnarray}
W_E=(W_{E,S})^{f_w}(W_{E,BH})^{1-f_w},
\end{eqnarray}
with weight
\begin{eqnarray}
f_w=\frac{1}{2}\Big\{1+\tanh\Big[-2.5\log_{10}\Big(\frac{E}{E_m}\Big)\Big]\Big\},
\end{eqnarray}
where $E_m$ is the energy where the distance between $W_{E,S}$ and $W_{E,BH}$ is minimum.

Taking $e=E_k/m_ec^2=\epsilon-1$, we have the following fitting
\begin{eqnarray}
W_{E,ei}=c_0\cdot f_{nr}\cdot f_{r}\cdot(1-f_z),
\end{eqnarray}
with
\begin{eqnarray}\nonumber
f_{nr}&=&1+c_{nr1}\Big\{1-\exp\Big[-\Big(c_{nr2}\frac{Z^2}{e}\Big)^{c_{nr4}}\Big]\Big\}-\\
    &&\Big(c_{nr1}+1-\frac{1}{c_0}\Big)\exp\Big[-\Big(\frac{e}{c_{nr3}Z^2}\Big)^{c_{nr5}}\Big],
\end{eqnarray}
\begin{eqnarray}\nonumber
f_{r}&=&1+1.069\frac{p^2}{\epsilon^2}-2.1939\frac{p^3}{\epsilon^3}+\\&&\Big[2.315\frac{p^2}{\epsilon^2}-1.5652\frac{p^4}{\epsilon^4}\Big]\epsilon\Big[\ln(p+\epsilon)-\frac{1}{3}\Big],
\end{eqnarray}
\begin{eqnarray}\label{eq:fzWE}
f_z=\frac{0.433Z\Big(10e/\sqrt{Z}\Big)^{-1.106}}{\exp\Big[2.8975\Big(e/\sqrt{{Z}}\Big)^{-0.256}\Big]-1}.
\end{eqnarray}
where
\begin{eqnarray*}
c_{nr}&=&[0.5165, ~   0.3246e-3, ~   0.2391e-4, ~   0.5907,  ~  0.3169].
\end{eqnarray*}

Fig.\ref{fig:WEeifit_xie} shows the result, with a total error less than 1\% for $Z=1-35$ [see Fig.\ref{fig:WEeifit_xie}b]. If $f_z$ is omitted, the max error for $Z=35$ is still less than 10\% [see Fig.\ref{fig:WEeifit_xie}b].

It is possible to find a better treatment of $f_z$. The correction of the $Z$ effects other than $f_{nr}$ and $f_r$ are not physically straightforward. So, we may choose to simplify it with a simple form. Another more simple fitting without $f_z$ can be accurate to within 1\% for $Z\leq7$ and within 2.5\% for $Z\leq35$ [see Fig.\ref{fig:WEeifit_xie}d]. The form is
\begin{eqnarray}
W_{E,ei}=c_0\cdot f_{nr}+f_{r},
\end{eqnarray}
with $f_r$ modified to
\begin{eqnarray}\nonumber
f_{r}&=&1.215\frac{p^2}{\epsilon^2}-2.3484\frac{p^3}{\epsilon^3}+\\&&\Big[2.344\frac{p^2}{\epsilon^2}-1.517\frac{p^4}{\epsilon^4}\Big]\epsilon\Big[\ln(p+\epsilon)-\frac{1}{3}\Big].
\end{eqnarray}
The above form could be sufficient for most usage in fusion plasma. 

The above fitting can be reduced to the extreme relativistic (ER) case\cite{Stickforth1961}, i.e.,
\begin{eqnarray}
W_{E}^{ER}(\epsilon)=6\epsilon\Big[\ln(2\epsilon)-\frac{1}{3}\Big].
\end{eqnarray}


\subsection{Fitting for e-e}
Define the normalized quantity in the center-of-mass system (C.M.S.) as
\begin{eqnarray}
Q_{cm}(\epsilon)=\frac{1}{\alpha r_e^2}\int_0^{k_{\rm{max}}}k_{\rm{cm}}\frac{d\sigma}{dk_{\rm{cm}}}dk_{\rm{cm}},
\end{eqnarray}
we can obtain the total e-e Bremsstrahlung power\cite{Haug1975b}:
\begin{eqnarray}\nonumber
W_{ee}&=&\frac{n_e^2m_ec^2\alpha r_e^2}{2}\int\int \frac{\epsilon_1+\epsilon_2}{\epsilon_1\epsilon_2}\sqrt{\frac{1}{2}[({\bm p}_1{\bm p}_2)-1]}\cdot\\&&f({\bm p}_1)f({\bm p}_2)Q_{cm}d^3p_1d^3p_2,
\end{eqnarray}
where the momenta and energies \(\epsilon_1,\epsilon_2,{\bm p}_1,{\bm p}_2\) are in any system \(S\). We have
\begin{eqnarray*}
p=|{\bm p}_1-{\bm p}_2|,~~\epsilon=\sqrt{p^2+1}.
\end{eqnarray*}
For Maxwellian electrons, the integral can reduce to\cite{Haug1989,Dermer1984}:
\begin{eqnarray}
W_{ee}&=&\frac{4n_e^2c\alpha r_e^2}{t[K_2(1/t)]^{2}}\int_1^{\infty}\epsilon^2 p^2K_2(2\epsilon/t)Q_{cm}(\epsilon)d\epsilon.
\end{eqnarray}
The \(Q_{cm}(\epsilon)\) can be obtained using the \(d\sigma/dk\) in \ref{sec:dsgmdkee} [for small energy, we use Eq.(\ref{eq:sgmnree}) to avoid numerical difficulty]. Hence, \(Q_{cm}(\epsilon)\) is the cross section which plays a central role in calculating the Bremsstrahlung power for arbitrary electron distributions, similar to the calculation of fusion reactivity\cite{Xie2023}. Haug\cite{Haug1975b,Haug1989} provided two fittings for \(Q_{cm}\). However, we find that the maximum error to the exact integral can be around 5\%, especially at \(\epsilon\sim100\) keV, as shown in Fig.\ref{fig:cmpQcm}. Thus, we provide a more accurate fitting here.

The extreme relativistic (ER) case (\(t\gg1\)) is\cite{Haug1975b}:
\begin{eqnarray}
Q_{cm}^{ER}(\epsilon)=16\epsilon\Big[\ln(2\epsilon)-\frac{1}{6}\Big],
\end{eqnarray}
and the nonrelativistic (\(t\ll1\)) case without Elwert factor is:
\begin{eqnarray}
Q_{cm}^{NR}(\epsilon)=8p^2=8(\epsilon^2-1).
\end{eqnarray}
Haug combined these two fits to cover all ranges\cite{Haug1975b}:
\begin{eqnarray}
Q_{cm}(\epsilon)\simeq8\frac{p^2}{\epsilon}\Big[1-\frac{4}{3}\frac{p}{\epsilon}+\frac{2}{3}\Big(2+\frac{p^2}{\epsilon^2}\Big)\ln(\epsilon+p)\Big],
\end{eqnarray}
which is then used to calculate the Maxwellian plasma Gaunt factor.

The below new but still simple fitting is found to fit the exact \(Q_{cm}\) for both with and without Elwert correction cases:
\begin{eqnarray}\label{eq:Qcmfit}\nonumber
Q_{cm}(\epsilon)&\simeq&8\frac{p^2}{\epsilon}F_{ee}\Big\{1.007+2.11\frac{p}{\epsilon}-3.45\frac{p^2}{\epsilon^2} +\\&&
2\Big[1+2.12\Big(\frac{p}{\epsilon}-1\Big)\Big]\ln(\epsilon+p)\Big\},
\end{eqnarray}
with
\begin{eqnarray}
F_{ee}(\epsilon)=\frac{1}{2}\Big(\tanh\Big\{0.62\Big[\log_{10}(\epsilon-1)+4.96\Big]\Big\}+1\Big).
\end{eqnarray}
For the with Elwert case at \(E_k/m_ec^2\geq10^{-5}\) and without Elwert case (\(F_{ee}=1\)), the fitting can agree with the exact value with an error of less than 1\%. Results are shown in Fig.\ref{fig:cmpQcm}. It is possible to improve the accuracy by including more terms.

\section{Summary and Conclusion}\label{sec:summ}

This work provides more accurate analytical formulas for Bremsstrahlung radiation power calculation in fusion plasmas, with errors less than 1\% for a wide energy range and high $Z$ (e.g., $Z=30$). For Maxwellian plasmas, the results are demonstrated using the Gaunt factor. For arbitrary distributions of electrons, accurate fitting formulas for the cross section are provided. The present fitting addresses some limitations of previous fittings. For example, the most recognized fitting by Svensson\cite{Svensson1982} is discontinuous at $t=1$ and also inaccurate for $T_e\leq20$ keV. Most fittings are inaccurate for high $Z$ and could be incorrect for $t\to0$.

Here, we are only interested in the total power. To study the accurate spectrum, one can use the models in section \ref{sec:model}, or refer to Hoof\cite{Hoof2015,Chluba2020} for $e$-$i$, and Nozawa\cite{Nozawa2009} for $e$-$e$. Another widely used Bremsstrahlung study in fusion plasmas is for fast electron studies such as runaway electrons or wave heating electrons colliding with the background ions and electrons\cite{Peysson2008}. { 
Incorporating the Bremsstrahlung effect into simulation models is crucial for studying the intricate physics of high-energy density plasmas, particularly in particle simulation codes \cite{Sentoku1998}.}

One should also notice that, due to $g_{ei}$ depending on $Z$, the commonly used $Z_{\rm eff}=\sum_i n_iZ_i^2/n_e$ for $g=Z_{\rm eff}g_{ei}+g_{ee}$ may not be rigorous. For accurate calculations of the total power for multi-ion plasmas, the term $[Zg_{ei}(t,Z)+g_{ee}(t)]$ in Eq.(\ref{eq:ptotal}) should be replaced by $[\sum_i Z_i^2\frac{n_i}{n_e}g_{ei}(t,Z_i)+g_{ee}(t)]$.

It should be noted that here we only use the available theoretical models, which are believed to be accurate enough for practical usage. Even more accurate calculations (cf. \cite{Tseng1971}) could be extremely complicated due to the sum of Feynman-Dyson graphs \cite{Elwert1969}. Some experimental comparisons are also available (cf. \cite{Pratt1975}). For e-i, we have not solved the complicated EH model, which may improve the accuracy for high $Z$ and could be future work. Here, we only provide the fitting for electron energy but have not discussed the distribution function effects (cf. \cite{Eidmann1975}) in detail, which could be future work.

The source codes to obtain the results of this work are available at \url{https://github.com/hsxie/brem}, which could be used for further developing other fittings.

\ack
The discussions with Yang Li, Yue-jiang Shi, and Xian-li Huang are acknowledged.

\appendix
\section{More fittings of thermal averaged Gaunt factor}\label{sec:morefit}

We provide here some other fittings for applications of different situations.

\subsection{Simplified Polynominal fitting}\label{sec:fitpoly}
If we need less terms of the fitting. The following one can be used. Keep to $t\leq2$ ($\sim1$MeV), with accurate to $\leq2$\% for $Z\leq5$ and $\leq5$\% for $Z\leq30$. We have
\begin{eqnarray}\label{eq:geifit3}
g_{ei}(t,Z)=c_0(1+f_{nr})+f_{r},
\end{eqnarray}
where we still use Eq.(\ref{eq:geinr}) for $f_{nr}$, but replace $f_r$ with
\begin{eqnarray*}
f_{r}&=&c_0(0.50t+2.9t^2-2.6t^3+1.12t^4-0.19t^5).
\end{eqnarray*}
The extra $f_z$ term is omitted. 

And for e-e, the following fitting
\begin{eqnarray}\label{eq:geefit3}\nonumber
g_{ee}(t)&=&F_{ee}\frac{6\sqrt{3}}{\sqrt{2}\pi}t[1.0+0.757t+1.81t^2-3.355t^3\\
&&+1.96t^4-0.388t^5],
\end{eqnarray}
can have an error less than 2\% for $t\leq2$.
The Elwert factor $F_{ee}(t)$ can also be omitted if one does not need accurate value at low temperature ($t\leq0.01$).

\subsection{One fitting for all range}\label{sec:fitall}
Similar to the fitting of energy, we can also comnbine the NR and ER to one fitting for thermal average cases.

For e-i, replace the $f_r$ with the below
\begin{eqnarray}\label{eq:geifit2}\nonumber
f_r
&\simeq&1.4502\sqrt{\frac{t}{t+1}}-2.677\Big(\sqrt{\frac{t}{t+1}}\Big)^2+\\\nonumber
&&\frac{9}{8}\sqrt{\frac{6}{\pi}}\sqrt{t}\Big[\ln(2t+1)+\frac{3}{2}-C_E\Big]\Big\{1+3.0\Big(\sqrt{\frac{t}{t+1}}-1\Big) +\\
&&-0.9198\Big[\Big(\sqrt{\frac{t}{t+1}}\Big)^3-1\Big]\Big\},
\end{eqnarray}
can also agree with the accurate value for all the range to an error less than 1.0\%. Without $f_z$, the max error is less than 1\% for $Z\leq5$, and less than 5\% for $Z\leq35$. That is to say, keeping only $f_{nr}$ and $f_r$ can be sufficient for most usage.

For e-e, the below
\begin{eqnarray}\label{eq:geefit2}\nonumber
g_{ee}(t)
&\simeq&F_{ee}(t)\frac{6\sqrt{3}}{\pi\sqrt{2}}\frac{t}{\sqrt{t+1}}\Big\{1.0-0.106\Big(\sqrt{\frac{t}{t+0.7}}\Big)^2 +\\\nonumber
&&\frac{3\sqrt{\pi}}{4}\Big[0.295+
3.347\Big(\sqrt{\frac{t}{t+0.7}}\Big)^2-\\&&2.642\Big(\sqrt{\frac{t}{t+0.7}}\Big)^3\Big]\ln(2t+1)\Big\},
\end{eqnarray}
agree with the accurate value to an error less than 0.5\%.

\section{Cross section of e-e}\label{sec:dsgmdkee}
We use the e-e cross section $d\sigma/dk$ from Haug89\cite{Haug1989}, which is integral of solid angle in C.M.S. of the Eq.(A1) of Haug75a \cite{Haug1975a}
\begin{eqnarray}\nonumber
\frac{d\sigma}{dk} = \frac{\alpha r_0^2}{\epsilon p k} F_{ee}\Big\{\sqrt{\frac{p^2 - \epsilon k}{\epsilon^2 - \epsilon k}} \Big[ \frac{1}{p}\ln \Big( \epsilon + p \Big) \Big\{2\epsilon + \frac{1}{\epsilon} + (\epsilon - k)
 \Big( 1 +\\\nonumber \frac{4 - 2\epsilon k}{p^2} + \frac{1}{p^4} + \frac{k}{\epsilon p^2 + k}\Big) 
+ \frac{p^2}{p^2 - \epsilon k} \Big( 4k - \epsilon + \frac{2k}{p^2 - \epsilon k} \Big)\Big\} \\\nonumber
- (\epsilon - k) \Big( \frac{16}{3}\epsilon + 2\frac{2\epsilon - k}{p^2} + \frac{1}{\epsilon p^4} \Big) - 4k^2 - \frac{k}{\epsilon} - \frac{2p^2}{p^2 - \epsilon k}\Big]
+ \\\nonumber
\frac{k}{2p} \sqrt{p^2 - \epsilon k} \ln \frac{2p\sqrt{\epsilon^2 - \epsilon k}+k}{2p\sqrt{\epsilon^2 - \epsilon k}- k} \Big[ \frac{\epsilon}{\epsilon p^2 + k} - \frac{\epsilon^2 + p^2}{p^2 - \epsilon k} - \\\nonumber \frac{2\epsilon k}{(p^2 - \epsilon k)^2} \Big] -
 \frac{\epsilon^2 k^2}{2p \sqrt{(p^2 - \epsilon k) (\epsilon^2 - \epsilon k)}} \ln \frac{2\epsilon p + (\epsilon - p) k}{2\epsilon p - (\epsilon + p) k}
\cdot \Big(\\\nonumber 2 \frac{\epsilon - k}{p^2 - \epsilon k} + \frac{\epsilon^2 + 1}{\epsilon p^2 + k} \Big) +
 \Big( 18 - \frac{2k}{\epsilon} + \frac{1}{\epsilon^2} \Big) L_1 + \frac{L_3}{ep} \Big[\\\nonumber \frac{32}{3} \epsilon^3 (\epsilon - k) + 8 \epsilon^2 k^2 + k^2 - 14\epsilon^2 - \frac{14}{3} p^2 +\\\nonumber
 \frac{19}{3}\epsilon k 
- \frac{k}{\epsilon} - \frac{k}{p^2} (\epsilon - k) - \frac{k}{2\epsilon p^4}-\frac{\epsilon^2+p^2}{ p^2 - \epsilon k} +\\\nonumber
 \frac{1}{p} \ln (\epsilon + p) \Big\{16\epsilon p^2 - 10p^2 k - 4\epsilon - k + \frac{11}{2\epsilon} + \frac{4k - \epsilon}{p^2} + \frac{k}{p^4} \\\nonumber
+ \frac{3(\epsilon^2 + p^2) - 4\epsilon^3 k}{2(p^2 - \epsilon k)} k + \frac{p^2}{2}\frac{\epsilon + k}{ (p^2 - \epsilon k)^2} - 2\frac{(\epsilon^2 + p^2)^3}{\epsilon^2 k}\Big\}+ \\\nonumber
 \frac{1}{p} \ln \frac{2p^2 - (\epsilon - p) k}{2p^2 - (\epsilon + p) k} \Big\{ 2\epsilon^3 - 4\epsilon + k + \frac{3}{4\epsilon} \\\nonumber
+ \frac{1}{p^2 - \epsilon k} \Big( \epsilon^3 k^2 - \frac{3}{4} k (\epsilon^2 + p^2) - \frac{p^2}{4} \frac{\epsilon + k}{ p^2 - \epsilon k} \Big)\Big\} \Big] +\\\nonumber
 \frac{2L_1 L_3}{p \sqrt{(\epsilon^2 - \epsilon k) (p^2 - \epsilon k)}} \Big[ 6p^2 k - 16\epsilon p^2 - \epsilon k^2 - 2\epsilon - \frac{3}{\epsilon}+\\\nonumber
\frac{(\epsilon^2 + p^2)^3 }{\epsilon^2 k}  \Big]
+ \frac{k^2}{\sqrt{\epsilon^2 - \epsilon k}} \int_{-1}^{1} d(\cos\theta) \Big[ \frac{2L}{\sqrt{R_1}}
 \Big\{\frac{2 - \epsilon^2}{\kappa_2} + \frac{2}{\kappa_1^2} +\\\nonumber \frac{\epsilon p^2}{\kappa_1 \kappa_2}(4\epsilon - k) + 
\frac{4p^2 - 3\epsilon k + 1/\kappa_1}{\kappa_1 (2p^2 - \kappa_2)} + \\\nonumber \frac{\epsilon}{\kappa_2 R_1} 
\Big[ \epsilon \kappa_1 - (\epsilon - k) \kappa_2 \Big] \Big( \frac{2p^2}{\kappa_1}+ \frac{\epsilon^2 + 1}{\epsilon^2 - \epsilon k} \Big)\Big\}+\\\nonumber
\frac{L_2}{W_2} \Big\{ 2 \frac{\epsilon^2 + p^2}{\kappa_2} - \frac{3}{2} + \frac{\epsilon^2 + p^2 - 2\epsilon k}{\epsilon k} \Big( \epsilon^2 - \epsilon k + \frac{\kappa_1}{4} \Big) + \\\nonumber
\frac{p^2}{2p^2 - \kappa_2} + \frac{1}{\kappa_1 (2p^2 - \kappa_2)} \Big[ (\epsilon^2 + p^2 -\\\nonumber 2\epsilon k)        \{ (\epsilon^2 + p^2 - \epsilon k) \kappa_2 - (\epsilon^2 + p^2) \} \\\nonumber
- \kappa_2 - \frac{\kappa_2^2}{2} \Big] + \frac{1}{4\epsilon k (2p^2 - \kappa_2)} \Big[ 2(\epsilon^2 + p^2 - 2\epsilon k) \Big\{\epsilon k + 1 \\\nonumber
- 4p^2 (p^2 - \epsilon k)\Big\} + \kappa_1^2 - 3\kappa_1 + (\epsilon^2 + p^2) \kappa_2 \Big]\Big\}\\\nonumber
- \frac{L_4}{W_4} \Big\{ 1 - \frac{\epsilon k}{2p^2} - \frac{1}{4p^4} + \frac{1}{\kappa_1 \kappa_2}(1 - \kappa_1 - \kappa_1^2) +\\ \frac{\epsilon^2 + p^2 - 2\epsilon k}{\kappa_1 \kappa_2} \Big[ 4p^2 (p^2 - \epsilon k) + \epsilon^2 \Big( 2k^2 - 2 + \frac{\epsilon k}{p^2} \Big) \Big] \Big\} \Big] \Big\},
\end{eqnarray}
where
\begin{eqnarray*}
\kappa_1 = k (\epsilon - p \cos\theta), \\
\kappa_2 = k (\epsilon + p \cos\theta),\\
L_1 = \ln \Big( \sqrt{\epsilon^2 - \epsilon k} + \sqrt{p^2 - \epsilon k} \Big), \\
L_2 = \ln \Big\{ 1 + 2\frac{\epsilon - k}{k \kappa_1} \Big[ (p^2 - \epsilon k) \kappa_2 + \sqrt{p^2 - \epsilon k} {W}_2 \Big] \Big\},\\
L_3 = \ln \frac{\Big( \epsilon \sqrt{p^2 - \epsilon k} + p \sqrt{\epsilon^2 - \epsilon k} \Big)^2}{\epsilon k}, \\
L_4 = \ln \Big\{ 1 + 2 \frac{\epsilon^2 - \epsilon k}{\kappa_1 \kappa_2} \Big[ 4p^2 (p^2 - \epsilon k) + \sqrt{p^2 - \epsilon k} {W}_4 \Big] \Big\},\\
L = \ln \Big\{ \frac{\sqrt{\epsilon^2 - \epsilon k}}{\kappa_1} \Big[ 2p^2 - \kappa_2 + \sqrt{(p^2 - \epsilon k)R_1}  \Big] \Big\}, \\
R_1 = 4 (p^2 - \epsilon k) + 4\kappa_1 + \frac{\kappa_1^2}{\epsilon^2 - \epsilon k},\\
{W}_2 = \sqrt{\frac{\epsilon \kappa_2}{\epsilon - k} \Big[ 2k^2 + (p^2 + k^2 - 2\epsilon k) \kappa_2 \Big]}, \\
{W}_4 = 2p \sqrt{4p^2 (p^2 - \epsilon k) + \frac{\kappa_1 \kappa_2}{\epsilon^2 - \epsilon k}},
\end{eqnarray*}
and Elwert correction factor
\begin{eqnarray}
F_{ee}=\frac{a_2}{a_1}\frac{e^{2\pi a_1}-1}{e^{2\pi a_2}-1},\\
a_1=\frac{\epsilon^2+p^2}{2\epsilon p}\alpha,\\
a_2=\frac{\epsilon^2+p^2-2\epsilon k}{2\sqrt{(\epsilon^2-\epsilon k)(p^2-\epsilon k)}}\alpha,
\end{eqnarray}
and normalized relativistic energy $
\epsilon=E_k/(m_ec^2)+1$, momentum $p=\sqrt{\epsilon^2-1}$, and $E_k$ be the kinetic energy, $k=\frac{h\nu}{m_ec^2}\in[0,k_{max}]$ with $k_{max}=p^2/\epsilon$. Although the above equation is tendous, the numerical calculation of it is straghtforward.
We have checked it numerically with $F_{ee}=1$, which can agree to the non-relativistic (NR) and extreme relativistic (ER) cases \cite{Haug1975a,Maxon1967}
\begin{eqnarray}\label{eq:sgmnree}
\frac{d\sigma^{NR}}{dk}=\frac{4}{15}\frac{\alpha r_0^2}{ k}F\Big(\frac{k}{p^2}\Big),
\end{eqnarray}
\begin{eqnarray}\nonumber
F(y)&=&\Big[17-\frac{3y^2}{(2-y)^2}\Big]\sqrt{1-y}+\ln\Big(\frac{1}{\sqrt{y}}+\sqrt{\frac{1}{y}-1}\Big)\\
&&\cdot\Big[\frac{12(2-y)^4-7(2-y)^2y^2-3y^4}{(2-y)^3}\Big],
\end{eqnarray}
and
\begin{eqnarray}
\frac{d\sigma^{ER}}{dk}=\frac{8\alpha r_0^2}{\epsilon k}\Big[\frac{4}{3}(\epsilon-k)+\frac{k^2}{\epsilon}\Big]\cdot\Big\{\ln\Big[\frac{4\epsilon^2}{k}(\epsilon-k)\Big]-\frac{1}{2}\Big\}.
\end{eqnarray}



\end{document}